\documentclass[twocolumn,showpacs,preprintnumbers,amsmath,amssymb]{revtex4}
\usepackage{graphicx}% Include figure files
\usepackage{dcolumn}% Align table columns on decimal point
\usepackage{bm}% bold math

\begin{document}

%\preprint{APS/123-QED}

%%%===============================================
\title{Coexistence of Dirac-Cone States and Superconductivity in Iron Pnictide Ba(Fe$_{1-x}$Ru$_x$As)$_2$}
%%%---

\author{Y. Tanabe$^1$}
 \thanks{Corresponding author: youichi@sspns.phys.tohoku.ac.jp}

\author{K. K. Huynh$^2$}

\author{S. Heguri$^2$}

\author{G. Mu$^2$}

\author{T. Urata$^2$}

\author{J. Xu$^1$}

\author{R. Nouchi$^1$}

\author{N. Mitoma$^2$}

\author{K. Tanigaki$^{1, 2}$}
 \thanks{Corresponding author: tanigaki@sspns.phys.tohoku.ac.jp}

\affiliation{$^1$WPI-Advanced Institutes of Materials Research, Tohoku University, Aoba, Aramaki, Aoba-ku, Sendai, 980-8578, Japan}

\affiliation{$^2$Department of Physics, Graduate School of Science, Tohoku University, Aoba, Aramaki, Aoba-ku, Sendai, 980-8578, Japan}

\date{\today}% It is always \today, today,

%%%*******************************************************
\begin{abstract}
The Ru doping effect on the Dirac cone states is investigated in iron pnictide superconductors Ba(Fe$_{1-x}$Ru$_x$As)$_2$ using the transverse magnetoresistance (MR) measurements as a function of temperature. 
The linear development of MR against magnetic field $B$ is observed for $x$ = 0 - 0.244 at low temperatures below the antiferromagnetic transition.
The $B$-linear MR is interpreted in terms of the quantum limit of the Dirac cone states by using the model proposed by Abrikosov.
An intriguing evidence is shown that the Dirac cone state persists on the electronic phase diagram where the antiferromagnetism and the superconductivity coexist.
\end{abstract}

\maketitle

%%%%%%%%%%%%%%%%%%%%%%%%%%%%%%%%%%%%%%%%%%%%%%%%%%%%%%%%%%%%%%%%%%%%%%%%%%%%%%%
%\section{Introduction}
%%%=============================================================
%%% the coherent superconducting (SC) state overcoming the thermal fluctuations

%\section*{Introduction}
Despite a quite large energy scale for the pairing interaction attributed to the exchange interaction $J$ between Cu spins ($\sim$ 2000 K), the small superfluid density and short coherence length of high-$T_{\rm c}$ cuprates can cause bulk superconducting (SC) state to break down due to the phase fluctuation \cite{Uemura, Uemura2}.
In order to realize the bulk SC state at the temperature as high as the pairing energy, the coherency of the Cooper paired electronic states must be enhanced.
In the case of a two-dimensional Pb granular system, it has been shown that deposition of Ag on a Pb film enhances intergranular interactions, resulting in the immediate development of a bulk SC state \cite{Merchant}.
In high-$T_{\rm c}$ cuprates, a similar increase in the bulk $T_{\rm c}$ has been reported for a bilayer film comprised of underdoped La$_{2-x}$Sr$_x$CuO$_4$ coated with heavily overdoped metallic La$_{2-x}$Sr$_x$CuO$_4$ \cite{Yuli}.
It has been proposed that the superconducting domains, which occur in the Fermi liquid states despite they are in fermiologic or nonfermiologic, form one after another as the carrier concentration increases and that coherent bulk SC occurs via Josephson coupling or proximity effects.
Thus, Fermi surfaces with high mobility behind Cooper paired electronic states are essential to realize a robust bulk SC state.
From a physics viewpoint, the coexistence of highly mobile carriers and superconductivity is important for achieving high-$T_{\rm c}$ superconductors, and the recent discovery of Dirac cone states in the another high-$T_{\rm c}$ material iron pnictide superconductors provide a good research stage to study this idea.
The Dirac cone state is a novel electronic state with ideal massless fermion character.
It has been theoretically predicted to exist in iron pnictide superconductors via special band folding below the antiferromagnetic transition temperature \cite{Fukuyama, Rien, Morinari} and experimentally confirmed in Ba(FeAs)$_2$ \cite{Richard, dhva, Khuong}. 
There are many reports showing that antiferromagnetic phase survives in iron pnictide superconductors in the underdoped regime of the electronic phase diagram, where $T_{\rm c}$ increases with an increase in concentration of substituents \cite{AFSC-muSR3, AFSC-NMR1, AFSC-NMR2, Canfield2}.
It has been reported recently that an inhomogeneous microscopic phase occurs between the superconducting phase and the magnetically ordered one in the case of Ba$_{1-x}$K$_x$Fe$_2$As$_2$ \cite{AFSC-muSR3, AFSC-NMR1}.
In addition, the microscopic coexistence of antiferromagnetism and superconductivity in Ba(Fe$_{1-x}$Co$_x$As)$_2$ has been suggested \cite{AFSC-NMR2}.
Consequently, highly mobile carriers in the Dirac cone states behind the Cooper paired electronic states in parabolic bands are thought to enhance the coherency of the Cooper pairs via a scattering process (Fig. 1).

%%%===
\begin{figure}
% Here is how to import EPS art
\includegraphics[width=0.85\linewidth]{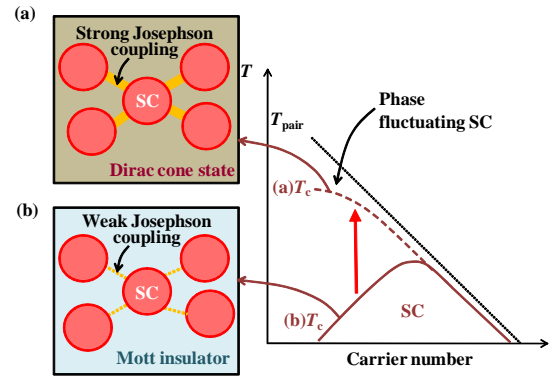}
\caption{(color online)  Concept of electronic phase diagram in high temperature superconductor (a) with Dirac cone states (b) with the Mott insulator. The strong Josephson coupling of SC islands due to the high mobility state of the Dirac cone immediately develops the bulk superconductivity in (a), while the bulk $T_{\rm c}$ is suppressed by the strong thermal fluctuation effect due to the weak coupling of SC islands in the insulating background in (b).}
\end{figure}
%%%===

%%% Ru doping  ------------------------
%%%===
We investigated this intriguing electronic states using a Ru substituted pnictide, Ba(Fe$_{1-x}$Ru$_x$As)$_2$, so that the Dirac cone states remained at high concentrations of the substituent.
Theoretically, it has been predicted that nonmagnetic impurities do not affect the Dirac cone states, whereas magnetic impurities destroy it \cite{Qi}.
Recent studies on the Nernst effect in Eu(Fe$_{1-x}$Co$_x$As)$_2$ and magnetotransport in Ba(Fe$_{1-x}$TM$_x$As)$_2$ (TM = Co, Ni, Cu) indicate that the influence of the Dirac fermion on electronic transport is greatly suppressed by substitution with magnetic impurities \cite{Matusiak, Fisher}.
Because the Ru 4d orbital is isoelectronic to the Fe 3d orbital, substitution with Ru should not disturb the Dirac cone state \cite{Canfield2, Canfield}.
Therefore, we thought that Ba(Fe$_{1-x}$Ru$_x$As)$_2$ was an ideal system for exploring the coexistence of superconductivity and Dirac cone states.
We showed that, by applying the transverse MR as a function of magnetic field ($B$) and temperature ($T$), the Dirac cone state persists in the underdoped regime of Ba(Fe$_{1-x}$Ru$_x$As)$_2$ and coexists with the superconductive phase.

%%% Experimental  %%%
%%%************************************************************
%\section{Experimental}
Single crystal of Ba(Fe$_{1-x}$Ru$_x$As)$_2$ were grown by the flux method using the FeAs flux.
Details were described in elsewhere \cite{Canfield}.
The qualities of single crystals were checked by the synchrotron X-ray diffraction measurements at the beam line BL02B2, SPring-8.
Electrical resistivity $\rho$ measurements were also carried out using the four-probe method to check the quality of the samples.
$B$ dependence of the in-plane transverse MR measurements were carried out using the 4 probe methods in -9 T $\leq$ $B$ $\leq$ 9 T at various fixed temperatures 2 K $\leq$ $T$ $\leq$ 150 K.

%%%=============================================================================
%\section{Results}
%%%===
\begin{table}
\begin{ruledtabular}
\begin{tabular}{cccccccc}
 $x$(Ru) & 0 & 0.0022 & 0.0051 & 0.102 & 0.150 & 0.190 & 0.244 \\
\hline
a(${\rm \AA}$) & 3.962 & 3.962 & 3.964 & 3.974 & 3.980 & 3.990 & 3.993 \\
c(${\rm \AA}$) & 13.02 & 13.01 & 12.99 & 12.96 & 12.92 & 12.90 & 12.86 \\
$T_{\rm s}$(K) & 137.2 & 131.6 & 122.5 & 99.1 & 83.9 & 63.7  & 46.5 \\
$T_{\rm c}$$^{\rm onset}$(K) &  -  &  -  &  -  & -  & 22.0 & 21.4 & 20.0 \\
$T_{\rm c}$(K) &  -  &  -  &  -  & -  & - & - & 11.5 \\

\end{tabular}
\end{ruledtabular}
\caption{The Ru concentration $x$, the lattice constant, the structure and the magnetic transition temperature $T_{\rm S}$, the onset and the middle of the superconducting transition temperature $T_{\rm c}$$^{\rm onset}$ and $T_{\rm c}$ for Ba(Fe$_{1-x}$Ru$_x$As)$_2$ single crystals.}
\end{table}
%%%-----------------------------------

The Ru concentration ($x$), the crystal lattice constants, the structure and magnetic transition temperature ($T_{\rm S}$), and the onset and the midpoint of the SC transition temperature ($T_{\rm c}$$^{\rm onset}$ and $T_{\rm c}$, respectively) for Ba(Fe$_{1-x}$Ru$_x$As)$_2$ single crystals are listed in Table 1.
The value of $x$ for Ba(Fe$_{1-x}$Ru$_x$As)$_2$ was determined by employing the relationship between the c-axis lattice constant and $x$, as previously reported.
The values of $T_{\rm S}$ were determined from the derivatives of the temperature dependences of $\rho$.
Details are described elsewhere \cite{Canfield}.

%%%***********************************************************************************
%%% Results and discussion
%%%************************************************************************************

%\section*{MR for Ba(Fe$_{1-x}$Ru$_x$As)$_2$}

%%%*****
The $B$ dependence of MR for Ba(Fe$_{1-x}$Ru$_x$As)$_2$ with $x$ = 0.190 at 20 K is shown in Fig. 2(a).
The evolution of MR as a function of $B$ is convex in the low- $B$ regime.
The $T$ dependence of the resistivity $\rho$ shows a large drop below 21.4 K due to the SC transition.
The convex curvature of MR suggests that a crossover occurs from an SC state at low $B$ to a normal state at high $B$.
In order to clarify the gradient of MR in the normal state, the $B$ dependence of the derivative of MR, dMR/d$B$, for $x$ = 0.190 is plotted in Fig. 2(b). A decrease in dMR/d$B$ was observed in a low $B$, and then it increased above 4 T.
In a high $B$, dMR/d$B$ became saturated.
In fact, the extrapolated line of dMR/d$B$ above $|\pm{\rm 4T}|$ shown in Fig.\,2(b) deviated from the ideal value of dMR/d$B$\,=\,0 under $B$ = 0 T, indicating that the MR cannot be described as MR $\propto$ $B^2$, which is the conventional MR behavior for metals in the low $B$ regime, but can be described as MR $\propto$ $B$ in the Dirac cone states in the high $B$ regime.

%%%===
\begin{figure}
% Here is how to import EPS art
\includegraphics[width=0.85\linewidth]{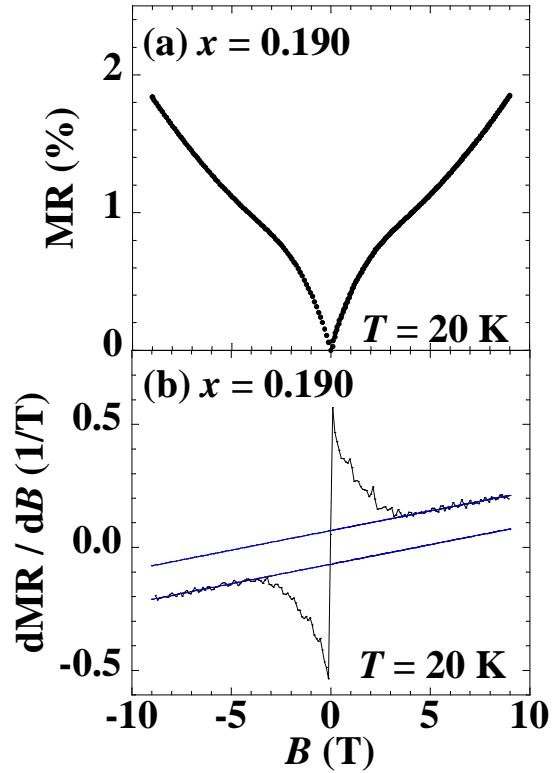}
\caption{(color online) (a) Magnetic field ($B$) dependence of the magnetoresistance (MR) for Ba(Fe$_{1-x}$Ru$_x$As)$_2$ with $x$ = 0.190 at 20 K. (b) The $B$ derivative of MR for $x$ = 0.190. The blue lines were fitted with dMR/d$B$ above $\mid$$B$$\mid$ = 4 T.}
\end{figure}
%%%===

The $B$ dependences of MR and dMR/d$B$ for Ba(Fe$_{1-x}$Ru$_x$As)$_2$ for various $x$ at 2 K are shown in Figs.3(a) and (b). 
For $x$ = 0.150-0.244, where the SC transition is observed below 20-22 K, the data at 22 K are shown in Figs. 3(a) and (b).
For $x$ = 0, MR developed linearly with $B$, and dMR/d$B$ saturated above 2 T.
The values of MR decreased with an increase in $x$.
As $x$ increased, dMR/d$B$ $\propto$ $B$ in a low $B$, and it saturated in a high $B$ for $x$ = 0-0.150.
We defined the critical magnetic field $B^*$ as the intercept point of the straight lines for dMR/d$B$ $\propto$ $B$ in the low $B$ regime and the one in the high $B$ regime where dMR/d$B$ saturates.
Above $x$ = 0.150, dMR/d$B$ deviated from a straight line with $B$ in the high $B$ regime, as shown for $x$ = 0.190 and 0.244 in the inset of Fig. 3(b). 

%%%===
\begin{figure}
\includegraphics[width=0.9\linewidth]{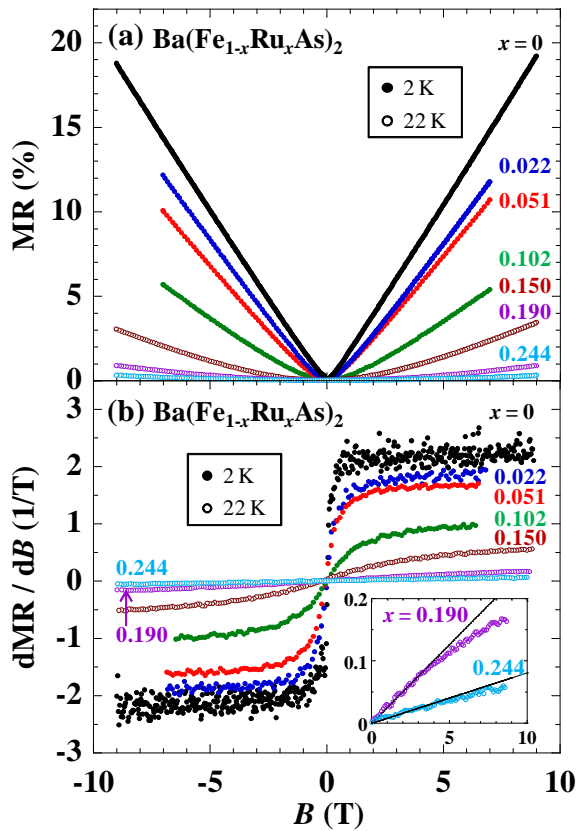}% Here is how to import EPS art
\caption{(color online) (a) Magnetic field ($B$) dependence of the magnetoresistance (MR) for Ba(Fe$_{1-x}$Ru$_x$As)$_2$ with $x$ = 0-0.150 at 2 K (close circle) and $x$ = 0.150-0.244 at 22 K (open circles). (b) $B$ derivative of MR for Ba(Fe$_{1-x}$Ru$_x$As)$_2$ with $x$ = 0-0.150 at 2 K (open circle) and $x$ = 0.150-0.244 at 22 K(closed circle). Inset of (b) shows a magnified plot of the $B$ derivative of MR for $x$ = 0.190 and 0.244. Solid lines represent linear fittings in the semiclassical region.}
\end{figure}
%%%===

%\section*{Interpretation of MR using Abrikosov model}

In order to analyze the data, we employed a theoretical model proposed by Abrikosov \cite{Abrikosov}.
The energy splitting between 0$^{\rm th}$ and 1$^{\rm st}$ Landau levels in the Dirac cone states is described in Eq. (1).
%%%===
\begin{eqnarray}
 \Delta_{\rm LL} &=& \displaystyle{\pm v_{\rm F}\sqrt{2{\rm e} {\rm \hbar} B}} \label{1}\\
 B^* &=& \displaystyle{(1/2{\rm e} {\rm \hbar} v_{\rm F}^2)(E_{\rm F} + k_{\rm B}T)^2} \label{2}
\end{eqnarray}
%%%}===
In the quantum limit, all carriers occupy the 0$^{\rm th}$ Landau level, and $\Delta_{\rm LL}$ becomes larger than both the Fermi energy ($E_{\rm F}$) and the thermal fluctuations at a finite temperature ($k_{\rm B}$$T$).
In this case, MR is not described using the semiclassical equation MR $\sim$ $B^2$ but using MR $\sim$ ($N_{\rm i}$/e$n_{\rm D}$$^2$)$B$, where $N_{\rm i}$ is the number of impurities and $n_{\rm D}$ is the number of carriers.
For a conventional parabolic band, a linearly $B$-dependent MR can be observed in the quantum limit.
However, the energy splitting of the Landau levels for a conventional parabolic band is described by $\Delta_{\rm LL}$ = e$\hbar$$B$/m$^*$ and the evolution of $\Delta_{\rm LL}$ with increasing $B$ is much slower than that for the Dirac cone.
Consequently, it is not possible to observe quantum limit behavior for the parabolic bands for Ba(Fe$_{1-x}$Ru$_x$As)$_2$ below 9 T.

%%%===
Figure 4(a) shows the temperature dependence of $B^*$ as a function of $T^2$.
For $x$ = 0.150, $B^*$ was estimated above 22 K because the temperature dependence of $\rho$ drops due to the SC transition below 22 K, as explained earlier.
The values of $B^*$ for $x$ = 0.190 and 0.244 were not estimated for accuracy because dMR/d$B$ was not sufficiently saturated.
The $B^*$ values increased monotonically with $T^2$ for $x$ = 0-0.150.
At $B$ = $B^*$, $\Delta_{\rm LL}$ can be expressed as the sum of $E_{\rm F}$ and $k_{\rm B}$$T$, as shown in Eq. (2).
The curves for each $x$ fitted with $B^*$ as a function of $T$ are shown as solid lines in Fig. 4(a),  and they agree with Eq. (2).

%%% Discussion
%%%===
%\section*{Discussion}
The linealy $B$-dependent MRs for Ba(Fe$_{1-x}$Ru$_x$As)$_2$ with $x$ = 0-0.244 are consistent with the quantum limit behavior of MR in a Dirac cone state \cite{Khuong}.
Moreover, the temperature dependence of the estimated $B^*$s for $x$ = 0-0.150 can be described using Eq. (2), which is based on the Abrikosov model, as described earlier.
Therefore, the Dirac cone states remain after substitution with Ru and are present in the electronic phase diagram where both antiferromagnetism and superconductivity occur at the same time.
We estimated both $E_{\rm F}$ and $v_{\rm F}$ using the value of $B^*$ and plotted them as a function of $x$ in Figs. 4(b) and (c).

%%%===
$E_{\rm F}$ increased with an increased in $x$, whereas $v_{\rm F}$ slightly decreased with an increase in $x$.
The results indicated that the Ru 4d orbitals caused a larger extension of the wave function than the Fe 3d orbitals did.
The on-site Coulomb repulsion (U) becomes smaller, and the bandwidth (W) becomes larger to modify the band reconstruction, influencing the band folding and, thus the Dirac cone states.
It is noted here that only the dominant terms from the Dirac cone states with
higher mobilities affect MR.
Density functional theory calculations suggest that substitution with Ru does not increase the number of carriers but does increase the bandwidth of the Fe 3d orbital via hybridization with the Ru 4d orbital, which has a larger spacial distribution of electrons \cite{Zhang}.
Experimentally, ARPES suggests that band renormalization occurs in the overdoped regime of Ba(Fe$_{1-x}$Ru$_x$As)$_2$ \cite{Brouet}, and thermoelectronic power measurements suggest that the Fermi surface topology changes in $x$ = 0.07 and 0.30 \cite{Canfield3}.
In other words, the increase in $E_{\rm F}$ that we observed above $x$ = 0.051 may be due to the change in the Fermi surface topology or band renormalization effects.
To the best of our knowledge, there are no reports on the vortex liquid phase as well as the SC fluctuations at temperatures well above the bulk $T_{\rm c}$ \cite{Matusiak, Nernst} for iron pnictide superconductors, although they are important characteristics of high-$T_{\rm c}$ cuprates \cite{Wang}.
In the present experiments, the Dirac cone states remain after substitution with Ru and coexist with superconductivity at low concentrations of Ru.
This suggests that the domains made by Cooper paired electrons are more coherent due to the high mobility of the Dirac cone states.
More detailed calculations are needed to explain the electronic phase diagram of Ba(Fe$_{1-x}$Ru$_x$As)$_2$ as well as $E_{\rm F}$ and $v_{\rm F}$ of the Dirac cone states determined in the present experiments.

%%%===
\begin{figure}
\includegraphics[width=1.0\linewidth]{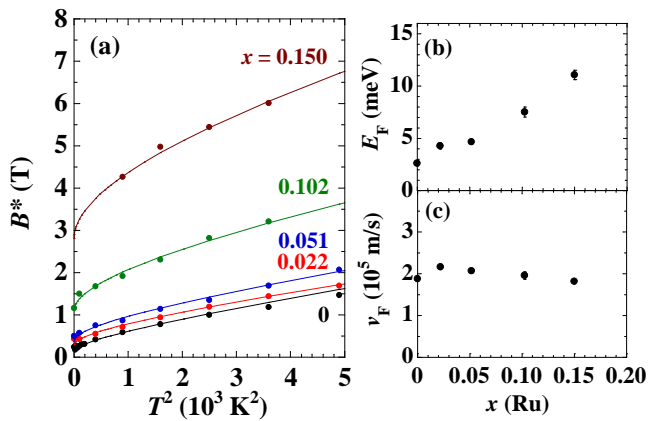}% Here is how to import EPS art
\caption{(color online) (a) $B^*$ vs. $T^2$ plot for Ba(Fe$_{1-x}$Ru$_x$As)$_2$ with $x$ = 0-0.150. The solid lines were fitted using $B^*$ = (1/2e$\hbar$$v_{\rm F}$$^2$)($E_{\rm F}$ + $k_{\rm B}$$T$)$^2$. The $x$ dependence of (b) $E_{\rm F}$ and (c) $v_{\rm F}$ for $x$ = 0-0.150.}
\end{figure}
%%%===

%%%%%%%%%%%%%%%%%%%%%%%%%%%%%%%%%%%%%%%%%%%%%%%%%%%%%%%%%%%%%%%%%%%%%%%%%%%%%%%
%\section*{Conclusion}
We investigated the effect of Ru doping in iron pnictide superconductor Ba(Fe$_{1-x}$Ru$_x$As)$_2$ on MR.
Linearly $B$-dependent MRs were observed below the structure and magnetic transition temperature ($T_{\rm S}$), and in the high $B$ regime, a normal state was observed for $x$ = 0-0.244, which is consistent with the quantum limit behavior of MR in the Dirac cone states.
$B^*$ values estimated for $x$ = 0-0.150 were explained in terms of the Landau level splitting for Dirac cone states.
Thus, we concluded that the Dirac cone states in Fe pnictides remain after substitution of Fe with Ru and are present in the electronic phase diagram where antiferromagnetism and superconductivity coexist.

The authors are grateful to P. Richard for useful comments.
The research was partially supported by Scientific Research on Priority Areas of New Materials Science using Regulated Nano Spaces, the Ministry of Education, Science, Sports and Culture, Grant in Aid for Science, and Technology of Japan.
The work was partly supported by the approval of the Japan Synchrotron Radiation Research Institute (JASRI).

%%%%%%%%%%%%%%%%%%%%%%%%%%%%%%%%%%%%%%%%%%%%%%%%%%%%%%%%%%%%%%%%%%%%%%%%%%%%%%%%%%

%*****************************************************************************************

\end{document}